\documentclass[traditabstract]{aa} 
\usepackage{amsmath}
\usepackage{txfonts} 
\usepackage{graphicx}
\usepackage{subfigure}
\usepackage{color}
\usepackage{bm}
\usepackage{natbib}
\usepackage{url}
\usepackage{lscape}
\usepackage{rotating}
\def\ltsima{$\; \buildrel < \over \sim \;$}
\def\simlt{\lower.5ex\hbox{\ltsima}}
\def\gtsima{$\; \buildrel > \over \sim \;$}
\def\simgt{\lower.5ex\hbox{\gtsima}}

\def\arcsec{\hbox{$^{\prime\prime}$}}

\begin{document} 
\title{CN Zeeman observations of the NGC 2264-C protocluster{\thanks{Fits file for the CN(1--0) map is only available in electronic form at the CDS via anonymous ftp to cdsarc.u-strasbg.fr (130.79.128.5) or via http://cdsweb.u-strasbg.fr/cgi-bin/qcat?J/A+A/}}$^{,}${\thanks{Based on observations carried out with the IRAM 30m Telescope. IRAM is supported by INSU/CNRS (France), MPG (Germany) and IGN (Spain).}}}
\author{Ana\"elle J. Maury\inst{1} 
\and Helmut Wiesemeyer\inst{2}
\and Clemens Thum\inst{3} 
}
\institute{ESO, Karl Schwarzschild Strasse 2, 85748 Garching bei M\"unchen, Germany
\and Max-Planck-Institut f\"ur Radioastronomie, Auf dem H\"ugel 69, 53121 Bonn, Germany
\and IRAM, 300 rue de la Piscine, 38406 St Martin d'H\`eres, France}
\date{Received 2012 March 1 / Accepted 2012 June 2} 
\abstract{From an observational point of view, the role of magnetic fields in star formation remains unclear, and two main theoretical scenarios have been proposed so far to regulate the star-formation processes. The first model assumes that turbulence in star-forming clumps plays a crucial role, and especially that protostellar outflow-driven turbulence is crucial to support cluster-forming clumps; while the second scenario is based on the consideration of a magnetically-supported clump. Previous studies of the NGC~2264-C protocluster indicate that, in addition to thermal pressure, some extra support might effectively act against the gravitational collapse of this cluster-forming clump. We previously showed that this extra support is not due to the numerous protostellar outflows, nor the enhanced turbulence in this protocluster. Here we present the results of the first polarimetric campaign dedicated to quantifying the magnetic support at work in the NGC~2264-C clump. 
Our Zeeman observations of the CN(1--0) hyperfine lines provide an upper limit to the magnetic field strength $B_{\rm{los}}$$\simlt$0.6 mG in the protocluster (projected along the line of sight). 
While these results do not provide sufficiently tight constraints to fully quantify the magnetic support at work in NGC~2264-C, they suggest that, within the uncertainties, the core could be either magnetically super or sub-critical, with the former being more likely.
}
\maketitle 

\section{Introduction} 

\subsection{Background: The role of magnetic fields in star formation}

It is well-established that stars form by gravitational collapse of molecular dense cores, themselves embedded in parsec-scale clumps contained in large molecular clouds (see, e.g. the review by \citealt{Larson03}). 
However, the conditions under which these dense cores form, collapse, and fragment remain a matter of debate. 
In particular, understanding the physics regulating the star-formation processes is crucial to, e.g, reconciling the values of star formation efficiencies observed on galactic scales, which are significantly lower than expected from purely gravitational collapse \citep{Zuckerman74, Krumholz07} predictions. 

Two main competing models have been proposed for driving and regulating the star-formation processes on the scales of star-forming clumps.
\\The first scenario argues that molecular clouds are intermittent structures in an interstellar medium dominated by turbulence (e.g. \citealt{Elmegreen00b}), and that turbulent motions prevent the clouds from collapsing in a freefall time \citep{MacLow04, Myers00a}. 
In this model, star-forming clumps undergo turbulent fragmentation \citep{Padoan02, Clark05}: self-gravitating pre-stellar condensations 
form as turbulence-generated density fluctuations, turbulence then dissipates rapidly, and the cores eventually collapse with little interaction with their surroundings. 
A somewhat different picture has been proposed in the case of clustered star formation. \citet{Nakamura07} 
argued that, owing to its short decay time \citep{MacLow98}, the interstellar turbulence that is initially present in a cluster-forming cloud is quickly replaced by turbulent motions generated by the numerous protostellar outflows. 
\citet{Li06} proposed that the protostellar outflow-driven turbulence dominates for most  of a protocluster's lifetime and acts to maintain the cluster-forming region close to overall virial equilibrium for several dynamical times, avoiding global free-fall collapse and reducing the local star-formation efficiency. 
\\The second model proposes that the magnetic field provides an efficient support against gravity and delays the star formation \citep{Shu87}, by 
forming self-gravitating dense clouds that are magnetically supported (e.g., \citealt{Mouschovias76, Ciolek01, Hennebelle08b}). 
The crucial parameter characterizing magnetic support of a dense cloud against its self-gravity is the ratio of the mass to magnetic flux M/$\Phi_{B}$. 
\citet{Mouschovias76} calculated the critical value of this mass-to-flux ratio using the virial theorem coupled to numerical calculations of a bidimensional cloud, and found that ($\rm{M}/\Phi_{B}$)$_{crit} \approx 0.13\times G^{-1/2}$, where $G$ is the gravitational constant. 
A cloud that has a mass-to-flux ratio larger than this critical value is said to be magnetically supercritical: it cannot be supported by magnetic pressure, and collapse can proceed on clump-sized scales, gravitationally bound dense cores eventually being formed. The cores progressively evolve towards higher degrees of central concentrations as the magnetic support is progressively lost through ambipolar diffusion. 
Otherwise, the cloud is subcritical: gravitational collapse is impeded because magnetic fields support the cloud against self-gravity. It is however possible that supersonic turbulence accelerates star formation in clouds that are initially subcritical, through enhanced ambipolar diffusion in shocks \citep{Li05}. 

One of the reasons for the ongoing debate about the respective roles of turbulence and magnetic fields in regulating star formation is the difficulty in accurately measuring the magnetic field in the dense environments typical of star-forming clumps.
On the observational side, only a handful of magnetic field strength estimates have been reported at densities $\simgt$10$^{5}$ cm$^{-3}$. While low-density HI structures (such as Giant Molecular Clouds) are found to be globally magnetically subcritical \citep{Heiles05}, most observations towards dense molecular cloud cores on sub-pc scales obtained so far suggest that they are approximately critical or supercritical \citep{Crutcher09}. 
Typically, magnetic fields detected in these dense molecular clouds have strengths 0.1 -- 1.5~mG \citep{Crutcher99a, Crutcher04, Falgarone08, Troland08, Heyer12}. 
We note, however, that some recent work based on the statistical analysis of the linear polarization suggests that some cloud cores (or regions of cloud cores) may be subcritical, such as the bowl of the Pipe Nebula \citep{Alves08, Franco10} or the massive star-forming region W51, on scales $\sim$1-2~pc \citep{Tang09, Koch12b}.

\subsection{Previous studies of the NGC~2264-C protocluster}

The NGC~2264-C cluster-forming clump is located in the Mon OB1 giant molecular cloud complex at d$\sim$ 800-900 pc \citep{Baxter09}.
Performing 30m and PdBI observations,  
\citet{Peretto06} carried out the first comprehensive millimeter continuum/line study of the protocluster.
Their 1.2~mm continuum mosaic of NGC~2264-C resolved the 
internal structure of the region, uncovering a total of 13 compact prestellar/protostellar cores.
Moreover, their line observations, combined with radiative transfer modeling, established the presence of large-scale collapse motions, converging onto the most massive core (C-MM3, near the center of NGC~2264-C ). 
Detailed comparison of these observations with numerical smoothed-particle hydrodynamics (SPH) simulations of the evolution of a 1000-M$_{\odot}$ Jeans-unstable isothermal clump \citep{Peretto07} suggested that NGC2264-C is an elongated clump collapsing/fragmenting along its long axis, and at a very early stage of global clump collapse ($\simlt 10^5$~yr after the start of dynamical contraction). 
A significant shortcoming of their SPH simulations, however, is that they only produce the observed level of clump fragmentation when the total mass of dense ($> 10^4\, \rm{cm}^{-3}$) gas in the model is a factor of $\sim 10$ lower than in the actual NGC~2264-C clump. This suggests the existence of forces that are not included in the hydrodynamical simulations and act as an extra support against gravity in NGC~2264-C, such as some support provided by feedback from protostellar outflows or magnetic fields.

To test the hypothesis of outflow-generated support suggested by \citet{Li06}, \citet{Maury09} used the heterodyne array HERA on the 30-m telescope to conduct a search for protostellar outflows, by mapping $^{12}$CO(2--1), $^{13}$CO(2--1), and C$^{18}$O(2--1) over the protocluster. They found a total of eleven outflow lobes originating from the dense protostellar cores forming in the protocluster. They carried out a quantitative study of the momentum flux injected by these outflows into the protocluster, and concluded that the network of outflows at work in NGC~2264-C fails to efficiently support the protocluster against global collapse.

The NGC~2264-C protocluster is therefore a well-studied star-forming clump, showing relatively simple global-collapse motions. 
To improve our understanding of this object but also use it as a case study for testing theories of clustered star formation, the magnetic field strength needs to be quantified in this dense clump. This should allow us to determine whether some magnetic support is responsible for the observed properties of the cluster-forming clump, and test the aforementioned theory of magnetized clouds.
In this paper, we present the results of the first polarimetric campaign dedicated to estimating the magnetic field strength in the NGC~2264-C cluster-forming clump. 

\subsection{CN Zeeman effect}

If a medium is permeated by a magnetic field $B$, the energy levels of the molecules and atoms split, owing to the interaction between the magnetic field and the magnetic dipole moment associated with their orbital angular momentum. 
The splitting of the energy levels then results in the splitting of the spectral lines into several separate polarized components: this effect, which was observed for the first time by \citet{Zeeman1897}, is the so-called Zeeman effect.

The Zeeman effect remains the only direct method for measuring magnetic field strengths in dense molecular clouds \citep{Crutcher07a, Crutcher09}, but the range of molecules that are at the same time abundant, paramagnetic, and accessible at millimeter wavelengths is rather small. 
As a tracer of high-density gas, the CN thermal lines probe the densest regions in molecular clouds.
Moreover, the N = 1$\rightarrow$ 0 transition of CN consists of nine hyperfine components (hereafter HFS), of which seven are strong components sensitive to the Zeeman effect. These HFS lines have different Zeeman splitting factors (see Table\,\ref{tab:CN_Zeeman} and \citealt{Turner75} for further references).
This property of the CN(1--0) emission lines is crucial for performing accurate Zeeman measurements probing the magnetic field, since the magnetically less-sensitive HFS (components Z2, Z3, and Z6 in Table~\ref{tab:CN_Zeeman}) can be used to correct for sidelobes and instrumental polarization effects (see \citealt{Crutcher96} for the detailed procedure). Since these instrumental effects can sometimes be stronger than the Zeeman signal produced by the magnetic field in the source itself \citep{Forbrich08}, the ability to make these corrections is crucial to distinguishing false detections from Zeeman signals that are truly magnetically-induced .

\begin{table}[!h]
\centering \par \caption{CN(1--0) strong hyperfine lines} 
\begin{tabular}{lcccccc|} 
\hline
\hline
{Line} & {Frequency} & {Relative} & {Zeeman splitting} & {$|$RI\,$\times$\,Z$|$ $^{a}$}\\ 
{} & {(GHz)} & {intensity RI} & {factor Z (Hz/$\mu$G)} & {}\\ 
\hline \\
{Z1} & {113.144} & {8} & {2.18} & {17.4}\\
{Z2} & {113.171} & {8} & {-0.31} & {2.5}\\ 
{Z3}& {113.191} & {10} & {0.62} & {6.2}\\ 
{Z4} & {113.488} & {10} & {2.18} & {21.8}\\
{Z5} & {113.491} & {27} & {0.56} & {15.1}\\
{Z6} & {113.500} & {8} & {0.62} & {5.0}\\
{Z7} & {113.509} & {8} & {1.62} & {13.0}\\
 \hline 
\end{tabular} 
\newline\noindent $^{a}$ $|$RI$\times$Z$|$ is the relative sensitivity to the magnetic field along the line of sight $B_{los}$.
\label{tab:CN_Zeeman} 
\end{table} 

\section{Observations and data analysis}

\subsection{CN mapping of the NGC~2264-C protocluster}

\begin{figure*}[ht]
\begin{center}
\includegraphics[width=0.75\linewidth,angle=0,trim=0cm 0cm 0cm 0cm,clip=true]{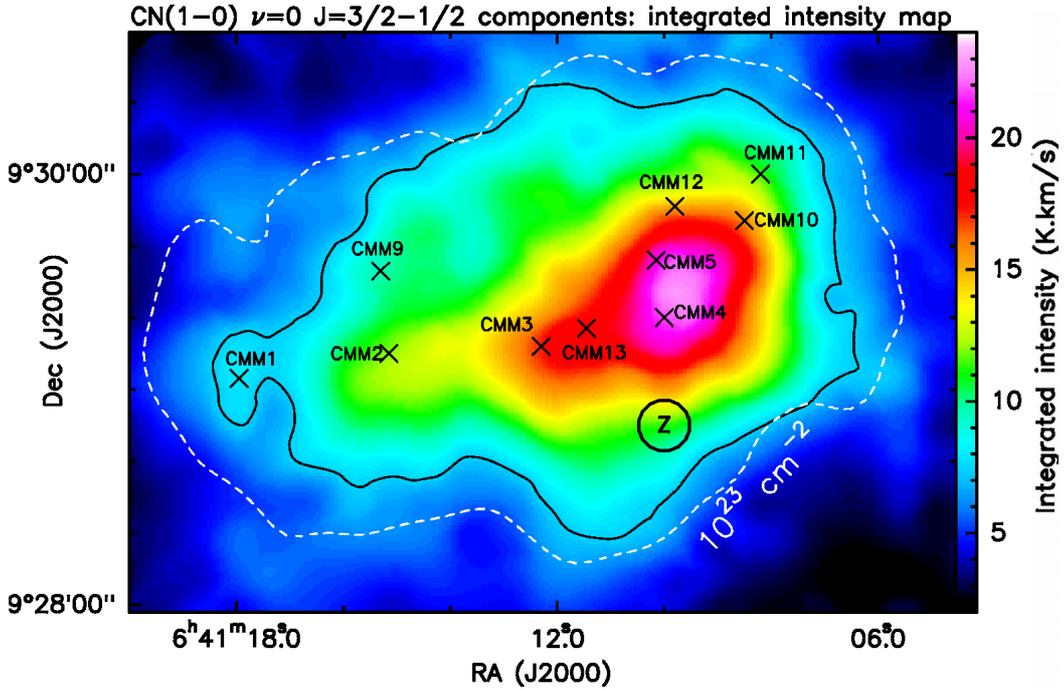}
\caption{Integrated intensity map of the CN ($N=1-0, J=3/2-1/2$) hyperfine components, obtained with EMIR in May 2009. The black contour represents the 3-$\sigma$ integrated intensity level (7.3~K.km.s$^{-1}$) in the map.
The crosses show the location of the millimeter dense cores mapped with MAMBO on the IRAM-30m by \citet{Peretto06}. The white dashed contour shows the value of column density $\sim$ 1$\times 10^{23}$~cm$^{-2}$, inferred from the MAMBO and P-ArT\'eMiS continuum maps \citep{Maury09}. 
The circle shows the location "Z" where we carried out further polarimetric observations to probe the Zeeman effect.
}
\label{fig:ngc2264_cn}
\end{center}
\end{figure*}

\citet{Peretto06} showed that the typical densities of the dense material in NGC~2264-C range from a few 10$^{5}\,\rm{cm}^{-3}$ to a few 10$^{6}\,\rm{cm}^{-3}$. With its critical density $n(H_{2})\sim 10^{5}$ cm$^{-3}$, CN is therefore an appropriate candidate molecule for conducting Zeeman observations in this region.

In preparation for future CN Zeeman observations, we first carried out an extensive mapping of the cluster-forming clump in the CN(1--0) line, to find positions where the line strength and profile would allow some Zeeman measurements.
These preliminary observations were carried out in May 2009 with the IRAM-30m telescope (FWHM of 23$\arcsec$ at 113\,GHz). 
We mapped a region of $5\arcmin \times 3\arcmin$ centered on C-MM3 (see Fig.~\ref{fig:ngc2264_cn}), using the on-the-fly mode. 
The Eight MIxer Receiver (EMIR) frontends were tuned to 113.3\,GHz, thus allowing us to observe simultaneously both the first (Z1, Z2, Z3) and the second (Z4 to Z7) group of CN(1--0) hyperfine components. The use of the VErsatile SPectrometer Assembly (VESPA) backends resulted in high spectral resolution (80~kHz i.e 0.1 km/s) maps for each group of HFS components.
We used simultaneously a third VESPA backend to cover a wide spectral window of 640\,MHz, with a lower spectral resolution of $\sim$0.8\,km.s$^{-1}$. 
With 12 hours of total observing time, we obtained root mean square (rms) noises in the final maps of $\sim$200~mK per 0.8\,km.s$^{-1}$-channel. 
Fig.\,\ref{fig:ngc2264_cn} shows the map of the CN(1--0) emission, integrated over the second group of HFS components (between 113.48 GHz and 113.52 GHz, transitions with $J=3/2-1/2$). The CN emission is strong towards the protocluster, and closely follows the dust continuum emission.

The spectra obtained towards the high density central region of the protocluster show double-peaked profiles, which could be due either to self-absorption, infall motions or the presence of two velocity components. To be able to distinguish between these three scenarios, we would need to perform observations in another optically thin, single-peaked transition of CN. We note, however, that earlier observations \citep{Peretto06, Peretto07} using other molecular tracers of dense gas have discovered the presence of large-scale infall motions in the protocluster. This strongly suggests that the double peak spectra are due to infall motions, but does not exclude other scenarios.

\begin{figure*}[ht]
\begin{center}
\includegraphics[width=0.75\linewidth,angle=0,trim=0cm 0cm 0cm 0cm,clip=true]{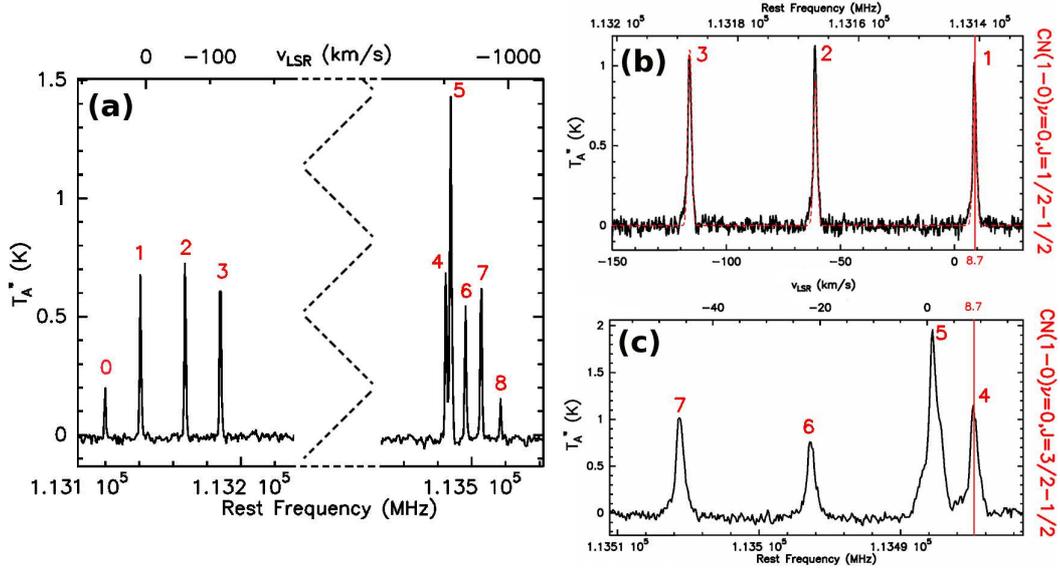}
\caption{CN(1-0) spectra from the "Z" position used for Zeeman observations, highlighted in the CN(1-0) integrated intensity map here-above. 
(a) Broad-band spectrum obtained at low spectral resolution (0.8\,km.s$^{-1}$ channel), covering the complete extent of all CN hyperfine components. (b) VESPA spectrum obtained towards the CN(1-0) triplet (J=1/2--1/2) at high spectral resolution (0.1\,km.s$^{-1}$ per channel originally, here smoothed to a 0.2\,km.s$^{-1}$ channel), towards the Z position. A fit of the hyperfine structure (red dashed spectrum) indicates that the systemic velocity is v$_{LSR} = 8.7 \pm 0.4$ km/s. (c) A VESPA spectrum obtained towards the CN(1-0) quadruplet (J=3/2--1/2) at high spectral resolution (0.1\,km.s$^{-1}$ per channel originally, here smoothed to a 0.2\,km.s$^{-1}$ channel), towards the Z position.}
\label{fig:Z_CN_spec}
\end{center}
\end{figure*}

In the external parts of the protocluster, the CN emission is well-represented by single-peaked Gaussian line profiles, which are indicative of an optically thin emission. Moreover, the spectra obtained in these external parts (see Fig.\,\ref{fig:Z_CN_spec}) show that although the measured relative intensities of the seven HFS components vary slightly from their optically thin LTE ratios (reported in Table\,\ref{tab:CN_Zeeman}), the variations are not large, thus suggesting that the CN emission is optically thin towards the outer parts of the clump.
Finally, we stress that the components "0" and "8" shown in Fig.\,\ref{fig:Z_CN_spec} are respectively the CN hyperfine components with quantum numbers (J=1/2-1/2, F=1/2-1/2) and (J=3/2-1/2, F=1/2-3/2), not used in the Zeeman analysis because their intensities are too weak.

\subsection{Polarimetric observations} 

Following \citet{Falgarone08}, we estimate that CN(1--0) lines stronger than $\sim$1K enable CN Zeeman observations sensitive to fields $B_{\rm{los}}$$\sim$0.2~mG within a moderate amount ($\simlt$40~hours) of telescope time, using the XPOL polarimeter on the IRAM-30m telescope \citep{Thum08}. 
The CN emission has to be close to optically thin, however, to avoid confusing the effects of optical thickness in the beam with a Zeeman signal. 
The "Z" position ($\alpha_{\rm{J2000}} =06^{\rm{h}}41^{\rm{m}}10.0^{\rm{s}},\, \delta_{\rm{J2000}} =09^{\circ}28\arcmin50.00\arcsec$) highlighted in Fig.\,\ref{fig:ngc2264_cn} is located near the center of the protocluster, and shows a strong non self-absorbed CN(1-0) line ($T_{\rm A}^* \sim 2$~K in the strongest hyperfine component, see Fig.~\ref{fig:Z_CN_spec}c). It was therefore chosen to perform deep polarimetric observations probing the Zeeman effect in the CN lines. 

The polarimetric observations with XPOL consist in using two orthogonally polarized 3-mm heterodyne receivers to obtain both the auto-correlation of each receiver (horizontal H and vertical V spectra) and the polarization cross-products (real and imaginary parts of the cross-correlation spectra). 
We used the 3mm receivers of EMIR associated with the VESPA backends in the polarimetry mode to carry out position-switch observations of the Z position, using an off position located at (-160$\arcsec$,-30$\arcsec$), checked to be clean of CN(1--0) emission.
The VESPA correlator was split into two separate parts, both with 80~kHz channel spacings and tuned to obtain simultaneous spectra of the two groups of CN(1--0) HFS components.
The spectra were first converted to a temperature scale by applying the standard calibration procedure for spectral observations (ambient load, cold load, and sky calibration). The correction for the phase difference between the receivers was done by using a polarization grid to observe the cold load during the calibration procedure (see \citealt{Thum08} for further details), resulting in a phase accuracy of $\simlt$0.1$^{\circ}$ rms. We checked the phase drift during our observations, estimating it to be less than 0.5$^{\circ}$ per hour. Finally, we fitted and subtracted the baselines with a polynomial of order 1, simultaneously for the two VESPA parts tuned to the two groups of hyperfine components.

To obtain significant constraints on the magnetic support at work in NGC~2264-C, we aimed to achieve a sensitivity $\sigma_{\rm rms} (T_{\rm A}^*) = 0.5$~mK in the Stokes V spectrum (imaginary part of the cross-correlation between the two polarizations, see \S4), and were granted a total of  80~hours of IRAM/30-m telescope time to perform polarimetric observations at the Z position.
The observations took place over one week at the end of April 2010, in mixed weather conditions (water vapor ranging from 4\,mm to 12\,mm and system temperatures from 200\,K to 700\,K) that did not allow us to reach the desired rms, but to obtain nevertheless good quality data suitable for a first meaningful analysis, that we present here.

\begin{center}
\begin{figure*}[!ht]
\begin{center}
\includegraphics[width=0.90\textwidth,angle=0,trim=0cm 0cm 0cm 0cm,clip=true]{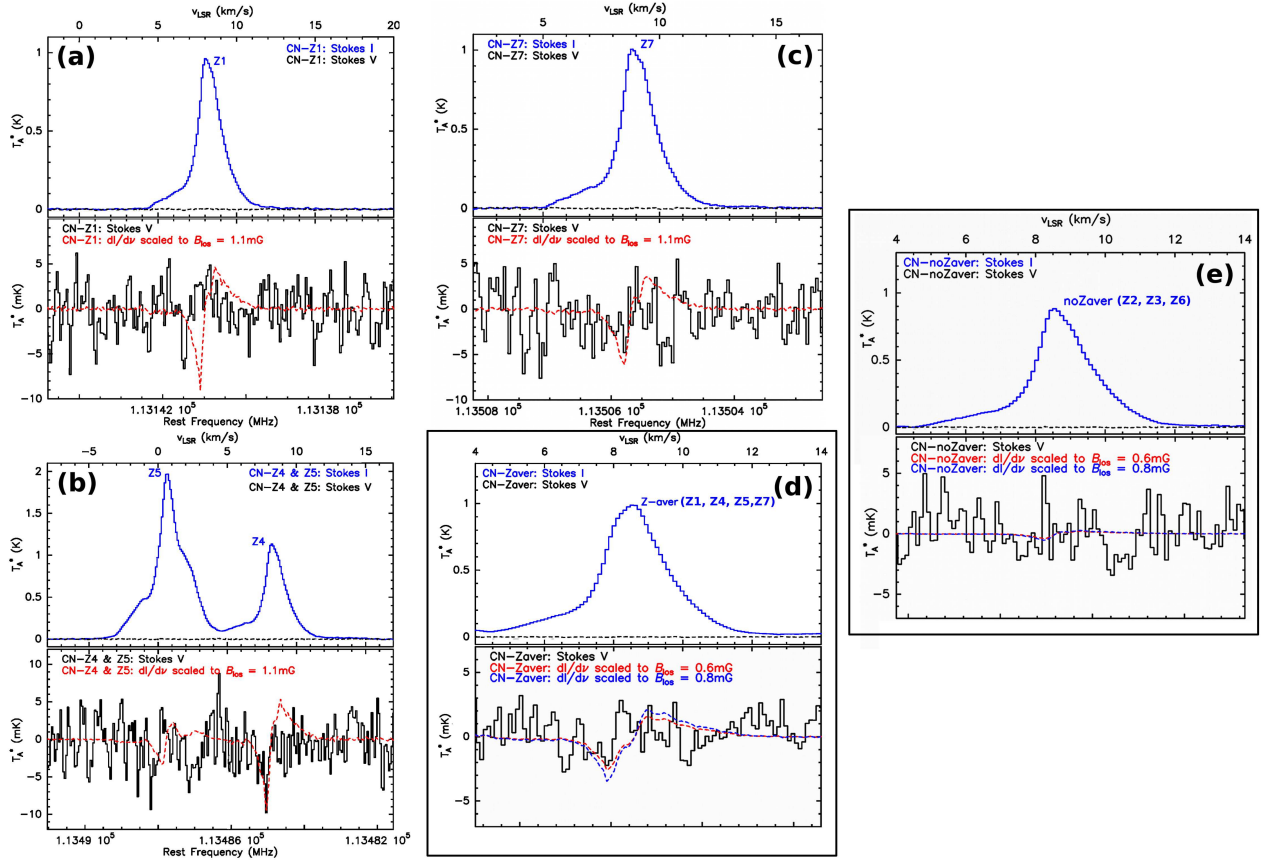}
\caption{(a) to (c) panels show the Stokes I (top panel) and Stokes V (bottom panel) cumulative spectra obtained for each of the Zeeman-sensitive CN(1--0) hyperfine components: component Z1 in (a), components Z4 and Z5 in (b), and component Z7 in (c). Overplotted on each Stokes V spectrum is the measurement of $dI/d\nu$ computed from the Stokes I spectrum, and scaled to the upper-limit magnetic field $B_{\rm{los}} \sim$1.1\,mG that can fit to each individual hyperfine Stokes V spectrum. The typical rms noise level in these spectra is $\sim$2.5\,mK. 
The panel (d) shows the average Stokes I (top) and Stokes V (bottom) spectra, using the four components that are most sensitive to the Zeeman effect (Z1, Z4, Z5, and Z7) weighted by their respective Zeeman sensitivity. The dashed colored lines overplotted in the lower panel show the $dI/d\nu$ computed from the average Stokes I spectrum and scaled to match the 3-$\sigma$ rms noise level in the Stokes V spectrum. The rms noise level in the average Stokes V spectrum is $\sigma_{\rm rms} (T_{\rm A}^*)$$\sim$1.3\,mK, leading to an upper limit on the magnetic field strength along the line of sight of ${B}_{\rm{los}} \simlt$ 0.6\,mG. The rightmost panel (e) shows the average Stokes I (top) and Stokes V (bottom) spectra, using the three components least sensitive to the Zeeman effect (Z2, Z3 and Z6) weighted by their respective Zeeman sensitivity. These components are mainly sensitive to the effects of instrumental polarization, and their spectra do not contain any significant feature.
 }
 \label{fig:Z_Stokes}
 \end{center}
\end{figure*}
\end{center}

\section{Results: upper-limit on $B_{\rm{los}}$ in NGC~2264-C}

Our observations provide us with the Stokes parameters. The Stokes I spectrum is the sum of the horizontal and vertical auto-correlation spectra $I = H + V$, while the circular-polarization component Stokes V is the imaginary part of the cross-correlation spectrum. 
In the case of a magnetic field ${B}_{\rm{los}}$ that is stronger than our detection limit, a Zeeman effect should be detected as an S-shaped signal of the form $V (\nu) = B_{\rm{los}} \times Z(dI/d\nu)$ in the Stokes V spectra of the hyperfine CN lines that are sensitive to magnetic fields (components Z1, Z4, Z5, and Z7). The insensitive components Z2, Z3, and Z6 should feature only noise if no significant instrumental effects are present.

The cumulative Stokes I and Stokes V spectra obtained for the Zeeman sensitive CN(1--0) hyperfine components are shown in 
Fig.\,\ref{fig:Z_Stokes}a, b, and c. To improve the signal-to-noise ratio and therefore the detectability of the magnetic field, we computed the average (weighted by the sensitivity to the Zeeman effect) Stokes I and Stokes V spectra derived from components Z1, Z4, Z5, and Z7.
The average spectra are shown in Fig.\,\ref{fig:Z_Stokes}d. 
Our observations do not detect any Zeeman signature: the average Stokes V spectrum contains only noise, and no significant S-shaped signal is detected. We note that the average Stokes V spectrum compiled from the insensitive components Z2, Z3, and Z6 can be well-fitted by a single Gaussian noise component (see Fig.~\ref{fig:Z_Stokes}e), showing therefore that the instrumental effects are neglictible at the sensitivity reached by our observations.
To illustrate the signal that would be observed in the Stokes V spectrum if a sufficiently strong magnetic field were present, we have over-plotted, in Fig.~3d, the $dI/d\nu$ signal computed from the average Stokes I spectrum and scaled for magnetic field strengths at our sensitivity limit $B_{\rm{los}} \sim$0.6--0.8\,mG (respectively, levels of 3-$\sigma$ and 4-$\sigma$).

\section{Discussion}

It is usual to define the ratio of the observed to critical mass-to-flux ratios $\mu=[M/\Phi]_{\rm{obs}}/[M/\Phi]_{\rm{crit}}$ as the characteristic quantity measuring the stability of magnetized clouds: values of $\mu$ above 1 correspond to cases where the magnetic field cannot support the cloud against its self-gravity and gravitational collapse will proceed.
The theoretical study of \citet{Nakano78} determined $\mu$ $=$7.6 $\times$10$^{-24}$ $N(H_{2})/ |B|$, where $N(H_{2})$ is in cm$^{-2}$ and $|B|$ is the magnetic field strength expressed in mG.
Considering the mean column density of NGC~2264-C ($N_{\rm{H_{2}}}$$\sim$1.8$\times10^{23}$cm$^{-2}$), the equipartition of the magnetic and gravitational energies is reached for a total magnetic field strength $|B|\sim$1.4\,mG: such a magnetic field (or stronger) would be able to prevent 
the global gravitational collapse of the whole NGC~2264-C clump. However, a magnetic field strength of half this critical value ($\mu=2$) is enough to provide significant support, slow down the gravitational collapse, and might explain the motions observed by \citet{Peretto07}. 

To efficiently constrain the magnetic support in NGC~2264-C, we initially aimed to probe mass-to-flux ratios $\mu \simgt$ 2, corresponding to total magnetic field strengths $B \simlt$ 0.6 mG.
Assuming a statistical mean inclination of the magnetic vector $\sim$57$^{\circ}$ with the line of sight (or applying the mean statistical geometrical correction proposed by \citealt{Heiles05}), the line-of-sight component probed by the Zeeman effect $B_{\rm{los}}$ is half the value of the total magnetic field vector norm $|B|$. 
Therefore, we aimed to reach a typical sensitivity limit $3 \sigma_{B_{\rm{los}}}\simlt$0.3 mG, corresponding to sensitivities $3 \sigma_{\rm rms} (T_{\rm A}^*) \simlt 1.5$~mK in the average Zeeman-sensitive (components Z1, Z4, Z5, and Z7) Stokes V spectrum.

Owing to the variable weather conditions during our 30-m observations, we could only reach $3\sigma_{\rm rms} (T_{\rm A}^*) = 3.6$~mK in the average Stokes V spectrum (see Fig.\,\ref{fig:Z_Stokes}), i.e. our observations provide an upper-limit to the magnetic field strength projected along the line of sight of 3$\times \sigma_{\rm rms}(B_{\rm{los}})$$\sim$0.6 mG.
Correcting the magnetic vector for statistical geometrical effects \citep{Heiles05}, 
our result turns into a tentative upper-limit to the field strength of $B \simlt$ 2$\times 3\sigma_{\rm rms}(B_{\rm{los}})$$\sim$ 1.2~mG. 
This is marginally above the dividing line between the magnetically subcritical and supercritical cases ($\mu \sim$ 1.1), suggesting that the NGC~2264-C cloud might be magnetically supercritical. 

We note that, at the position chosen for our Zeeman measurements, the CN line intensity is subject to a slight gradient across the 23$\arcsec$ half-power beam width (HPBW) (see Fig.~\ref{fig:ngc2264_cn}), which we estimate to be $\sim$1--2~K, increasing along a south-north axis. 
This implies that the instrumental Stokes V will get a sign, depending on the sidelobe (the positive or negative one) where the emission is the strongest.
However, since this intensity gradient is similar on all velocity channels, it cannot mimic a Zeeman feature or artificially erase one: it is therefore unlikely that the CN intensity gradient could affect our Zeeman measurements.
In addition, the Z position is located at the border of the blue outflow lobe F2  and in the red outflow lobe F11 traced in the $^{12}$CO emission maps of \citet{Maury09}. We note that the CN profiles obtained in Stokes I might partly trace some outflowing gas. This would explain the rather structured velocity profiles of the CN lines seen for example in Figs.~2 and~3 which both have a prominent blue wing and a fainter red shoulder seen in the strongest components.
These velocity features do not jeopardize our Zeeman analysis, however, since we search for Zeeman signatures near the emission maximum, i.e around the systemic velocity of the cloud where the emission is dominated by quiescent material and therefore outflowing material should not significantly contaminate neither the observed Stokes I nor Stokes V.

However, we stress that several factors can lead to the non-detection of a Zeeman signal even in the case of strong magnetic field. 
First, unfavorable inclination effects can result in a very faint line-of-sight component of the magnetic field, while the absolute strength of the magnetic vector could be significant and even sufficient to affect the clump dynamics on large scales. For example, \citet{Ostriker01} suggested that mean line-of-sight magnetic field strengths may vary widely across a projected cloud.
It is also possible that complex motions in the protocluster act to cancel or weaken any Zeeman signal in our observations of the CN(1--0) emission lines, obtained in a relatively large beam (23$\arcsec$ HPBW, corresponding to a projected distance of 0.1\,pc in the plane of the sky). 
For example, \citet{Li-HB11} recently showed that strong coupling between turbulent gas and the magnetic field can lead to complex magnetic field morphologies that are strongly dependent on the turbulence properties and the local density.

Deeper Zeeman observations allowing us to reach weaker levels of polarization (down to $\sim$$B_{\rm{los}} \simlt$ 0.1\,mG, similar to the magnetic field strength observed in the nearby clump NGC~2264-D by \citealt{Kwon11}) are needed to confirm our first results presented here. 
Additional observations probing the component of the magnetic vector in the plane of the sky, such as dust polarization observations in either the infrared \citep{Sugitani10, Chapman11} or the submillimeter regime \citep{Matthews02e, Koch12a} are crucially needed to definitively rule out the possibility of strong magnetic support in NGC~2264-C.

~\\
{\it{Acknowledgements:}} The research leading to these results has received funding from the European CommunityÕs Seventh Framework Programme (/FP7/2007-2013/) under grant agreement No. 229517.

 \bibliographystyle{aa}

\begin{thebibliography}{45}
\expandafter\ifx\csname natexlab\endcsname\relax\def\natexlab#1{#1}\fi

\bibitem[{{Alves} {et~al.}(2008){Alves}, {Franco}, \& {Girart}}]{Alves08}
{Alves}, F.~O., {Franco}, G.~A.~P., \& {Girart}, J.~M. 2008, \aap, 486, L13

\bibitem[{{Baxter} {et~al.}(2009){Baxter}, {Covey}, {Muench}, {F{\H
  u}r{\'e}sz}, {Rebull}, \& {Szentgyorgyi}}]{Baxter09}
{Baxter}, E.~J., {Covey}, K.~R., {Muench}, A.~A., {et~al.} 2009, \aj, 138, 963

\bibitem[{{Chapman} {et~al.}(2011){Chapman}, {Goldsmith}, {Pineda}, {Clemens},
  {Li}, \& {Kr{\v c}o}}]{Chapman11}
{Chapman}, N.~L., {Goldsmith}, P.~F., {Pineda}, J.~L., {et~al.} 2011, \apj,
  741, 21

\bibitem[{{Ciolek} \& {Basu}(2001)}]{Ciolek01}
{Ciolek}, G. \& {Basu}, S. 2001, \apj, 547, 272

\bibitem[{{Clark} \& {Bonnell}(2005)}]{Clark05}
{Clark}, P.~C. \& {Bonnell}, I.~A. 2005, \mnras, 361, 2

\bibitem[{{Crutcher} {et~al.}(2004){Crutcher}, {Nutter}, {Ward-Thompson}, \&
  {Kirk}}]{Crutcher04}
{Crutcher}, R., {Nutter}, D., {Ward-Thompson}, D., \& {Kirk}, J. 2004, \apj,
  600, 279

\bibitem[{{Crutcher} {et~al.}(1996){Crutcher}, {Troland}, {Lazareff}, \&
  {Kazes}}]{Crutcher96}
{Crutcher}, R., {Troland}, T., {Lazareff}, B., \& {Kazes}, I. 1996, \apj, 456,
  217

\bibitem[{{Crutcher} {et~al.}(1999){Crutcher}, {Troland}, {Lazareff},
  {Paubert}, \& {Kaz{\`e}s}}]{Crutcher99a}
{Crutcher}, R., {Troland}, T., {Lazareff}, B., {Paubert}, G., \& {Kaz{\`e}s},
  I. 1999, \apjl, 514, L121

\bibitem[{{Crutcher}(2007)}]{Crutcher07a}
{Crutcher}, R.~M. 2007, in EAS Publications Series, Vol.~23, EAS Publications
  Series, ed. {M.-A.~Miville-Desch{\^e}nes \& F.~Boulanger}, 37--54

\bibitem[{{Crutcher}(2009)}]{Crutcher09}
{Crutcher}, R.~M. 2009, in Revista Mexicana de Astronomia y Astrofisica
  Conference Series, Vol.~36, Revista Mexicana de Astronomia y Astrofisica
  Conference Series, 107--112

\bibitem[{{Elmegreen}(2000)}]{Elmegreen00b}
{Elmegreen}, B. 2000, \apj, 530, 277

\bibitem[{{Falgarone} {et~al.}(2008){Falgarone}, {Troland}, {Crutcher}, \&
  {Paubert}}]{Falgarone08}
{Falgarone}, E., {Troland}, T., {Crutcher}, R., \& {Paubert}, G. 2008, \aap,
  487, 247

\bibitem[{{Forbrich} {et~al.}(2008){Forbrich}, {Wiesemeyer}, {Thum},
  {Belloche}, \& {Menten}}]{Forbrich08}
{Forbrich}, J., {Wiesemeyer}, H., {Thum}, C., {Belloche}, A., \& {Menten},
  K.~M. 2008, \aap, 492, 757

\bibitem[{{Franco} {et~al.}(2010){Franco}, {Alves}, \& {Girart}}]{Franco10}
{Franco}, G.~A.~P., {Alves}, F.~O., \& {Girart}, J.~M. 2010, \apj, 723, 146

\bibitem[{{Heiles} \& {Crutcher}(2005)}]{Heiles05}
{Heiles}, C. \& {Crutcher}, R. 2005, in Lecture Notes in Physics, Berlin
  Springer Verlag, Vol. 664, Cosmic Magnetic Fields, ed. {R.~Wielebinski \&
  R.~Beck}, 137

\bibitem[{{Hennebelle} \& {Teyssier}(2008)}]{Hennebelle08b}
{Hennebelle}, P. \& {Teyssier}, R. 2008, \aap, 477, 25

\bibitem[{{Heyer} \& {Brunt}(2012)}]{Heyer12}
{Heyer}, M.~H. \& {Brunt}, C.~M. 2012, \mnras, 420, 1562

\bibitem[{{Koch} {et~al.}(2012{\natexlab{a}}){Koch}, {Tang}, \& {Ho}}]{Koch12a}
{Koch}, P.~M., {Tang}, Y.-W., \& {Ho}, P.~T.~P. 2012{\natexlab{a}}, \apj, 747,
  79

\bibitem[{{Koch} {et~al.}(2012{\natexlab{b}}){Koch}, {Tang}, \& {Ho}}]{Koch12b}
{Koch}, P.~M., {Tang}, Y.-W., \& {Ho}, P.~T.~P. 2012{\natexlab{b}}, \apj, 747,
  80

\bibitem[{{Krumholz} \& {Tan}(2007)}]{Krumholz07}
{Krumholz}, M.~R. \& {Tan}, J.~C. 2007, \apj, 654, 304

\bibitem[{{Kwon} {et~al.}(2011){Kwon}, {Tamura}, {Kandori}, {Kusakabe},
  {Hashimoto}, {Nakajima}, {Nakamura}, {Nagayama}, {Nagata}, {Hough}, {Werner},
  \& {Teixeira}}]{Kwon11}
{Kwon}, J., {Tamura}, M., {Kandori}, R., {et~al.} 2011, \apj, 741, 35

\bibitem[{{Larson}(2003)}]{Larson03}
{Larson}, R.~B. 2003, Reports on Progress in Physics, 66, 1651

\bibitem[{{Li} {et~al.}(2011){Li}, {Blundell}, {Hedden}, {Kawamura}, {Paine},
  \& {Tong}}]{Li-HB11}
{Li}, H.-B., {Blundell}, R., {Hedden}, A., {et~al.} 2011, \mnras, 411, 2067

\bibitem[{{Li} \& {Nakamura}(2006)}]{Li06}
{Li}, Z.-Y. \& {Nakamura}, F. 2006, \apjl, 640, L187

\bibitem[{{Mac Low} \& {Klessen}(2004)}]{MacLow04}
{Mac Low}, M.-M. \& {Klessen}, R. 2004, Reviews of Modern Physics, 76, 125

\bibitem[{{Mac Low} {et~al.}(1998){Mac Low}, {Klessen}, {Burkert}, \&
  {Smith}}]{MacLow98}
{Mac Low}, M.-M., {Klessen}, R.~S., {Burkert}, A., \& {Smith}, M.~D. 1998,
  Physical Review Letters, 80, 2754

\bibitem[{{Matthews} \& {Wilson}(2002)}]{Matthews02e}
{Matthews}, B.~C. \& {Wilson}, C.~D. 2002, \apj, 574, 822

\bibitem[{{Maury} {et~al.}(2009){Maury}, {Andr{\'e}}, \& {Li}}]{Maury09}
{Maury}, A., {Andr{\'e}}, P., \& {Li}, Z.-Y. 2009, \aap, 499, 175

\bibitem[{{Mouschovias} \& {Spitzer}(1976)}]{Mouschovias76}
{Mouschovias}, T. \& {Spitzer}, Jr., L. 1976, \apj, 210, 326

\bibitem[{{Myers}(2000)}]{Myers00a}
{Myers}, P. 2000, \apjl, 530, L119

\bibitem[{{Nakamura} \& {Li}(2005)}]{Li05}
{Nakamura}, F. \& {Li}, Z.-Y. 2005, \apj, 631, 411

\bibitem[{{Nakamura} \& {Li}(2007)}]{Nakamura07}
{Nakamura}, F. \& {Li}, Z.-Y. 2007, \apj, 662, 395

\bibitem[{{Nakano} \& {Nakamura}(1978)}]{Nakano78}
{Nakano}, T. \& {Nakamura}, T. 1978, \pasj, 30, 671

\bibitem[{{Ostriker} {et~al.}(2001){Ostriker}, {Stone}, \&
  {Gammie}}]{Ostriker01}
{Ostriker}, E.~C., {Stone}, J.~M., \& {Gammie}, C.~F. 2001, \apj, 546, 980

\bibitem[{{Padoan} \& {Nordlund}(2002)}]{Padoan02}
{Padoan}, P. \& {Nordlund}, {\AA}. 2002, \apj, 576, 870

\bibitem[{{Peretto} {et~al.}(2006){Peretto}, {Andr{\'e}}, \&
  {Belloche}}]{Peretto06}
{Peretto}, N., {Andr{\'e}}, P., \& {Belloche}, A. 2006, \aap, 445, 979

\bibitem[{{Peretto} {et~al.}(2007){Peretto}, {Hennebelle}, \&
  {Andr{\'e}}}]{Peretto07}
{Peretto}, N., {Hennebelle}, P., \& {Andr{\'e}}, P. 2007, \aap, 464, 983

\bibitem[{{Shu} {et~al.}(1987){Shu}, {Adams}, \& {Lizano}}]{Shu87}
{Shu}, F., {Adams}, F., \& {Lizano}, S. 1987, \araa, 25, 23

\bibitem[{{Sugitani} {et~al.}(2010){Sugitani}, {Nakamura}, {Tamura},
  {Watanabe}, {Kandori}, {Nishiyama}, {Kusakabe}, {Hashimoto}, {Nagata}, \&
  {Sato}}]{Sugitani10}
{Sugitani}, K., {Nakamura}, F., {Tamura}, M., {et~al.} 2010, \apj, 716, 299

\bibitem[{{Tang} {et~al.}(2009){Tang}, {Ho}, {Koch}, {Girart}, {Lai}, \&
  {Rao}}]{Tang09}
{Tang}, Y.-W., {Ho}, P.~T.~P., {Koch}, P.~M., {et~al.} 2009, \apj, 700, 251

\bibitem[{{Thum} {et~al.}(2008){Thum}, {Wiesemeyer}, {Paubert}, {Navarro}, \&
  {Morris}}]{Thum08}
{Thum}, C., {Wiesemeyer}, H., {Paubert}, G., {Navarro}, S., \& {Morris}, D.
  2008, \pasp, 120, 777

\bibitem[{{Troland} \& {Crutcher}(2008)}]{Troland08}
{Troland}, T. \& {Crutcher}, R. 2008, \apj, 680, 457

\bibitem[{{Turner} \& {Gammon}(1975)}]{Turner75}
{Turner}, B.~E. \& {Gammon}, R.~H. 1975, \apj, 198, 71

\bibitem[{{Zeeman}(1897)}]{Zeeman1897}
{Zeeman}, P. 1897, \apj, 5, 332

\bibitem[{{Zuckerman} \& {Evans}(1974)}]{Zuckerman74}
{Zuckerman}, B. \& {Evans}, II, N.~J. 1974, \apjl, 192, L149

\end{thebibliography}

\end{document}